\documentclass[a4paper]{jpconf}
\usepackage{graphicx}
\usepackage{amssymb}
\begin{document}

\newcommand{\lP}{\ell_{\rm P}}
\newcommand{\md}{{\mathrm d}}
\newcommand{\vt}{\vartheta}
\newcommand{\vp}{\varphi}
\newcommand{\uvec}[1]{\raisebox{-1.5mm}{$\stackrel{\textstyle #1}{\scriptscriptstyle\rightarrow}$}{}}

\title{Quantum gravity, space-time structure,\\ and cosmology}

\author{Martin Bojowald}

\address{The Pennsylvania State University,
University Park, PA, USA}

\ead{bojowald@gravity.psu.edu}

\begin{abstract}
  A set of diverse but mutually consistent results obtained in different
  settings has spawned a new view of loop quantum gravity and its physical
  implications, based on the interplay of operator calculations and effective
  theory: Quantum corrections modify, but do not destroy, space-time and the
  notion of covariance. Potentially observable effects much more promising
  than those of higher-curvature effective actions result; loop quantum
  gravity has turned into a falsifiable framework, with interesting
  ingredients for new cosmic world views. At Planckian densities, space-time
  disappears and is replaced by 4-dimensional space without evolution.
\end{abstract}

Can we expect quantum-gravity observations?  Even during inflation --- the
most energetic cosmological regime to which we have at least some indirect
access --- the Planck length $\ell_{\rm P}$ is far removed, by many orders of
magnitude, from the cosmological scale $\ell_{\cal H}=1/{\cal H}$, the inverse
Hubble parameter. Any higher-curvature action
\begin{equation}\label{HigherCurv}
 S_{\rm higher\: curvature}= \frac{1}{16\pi G} \int{\rm d}^4x \sqrt{-\det g}
 (\Lambda+R+ A \ell_{\rm P}^2R^2+\cdots)\,,
\end{equation}
often expected as the universal low-energy manifestation of quantum gravity,
is then all but identical to the Einstein--Hilbert action. No sizeable effects
of higher-order terms in the expansion should appear.  Quantum gravity might
be significant before inflation, at even higher energy densities, but then
most effects are likely washed away during inflation and afterwards.

There is a more optimistic alternative: quantum gravity might modify
space-time structure. An action covariant under the deformed symmetries of
modified space-time would not be of the standard higher-curvature form.  New
possibilities for quantum-gravity effects arise.  General features of
quantized gravity show that such a scenario is to be expected: Gravitational
dynamics is equivalent to space-time structure; evolving forward in time means
probing different parts of space-time, from different perspectives. Quantum
corrections in Hamiltonians, governing evolution of space-time and matter,
could then imply modifications of space-time structure.

\begin{figure}[h]
\includegraphics[width=14pc]{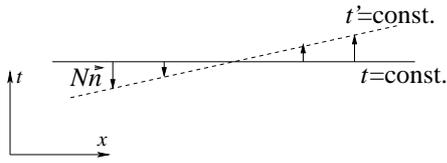}\hspace{2pc}%
\begin{minipage}[b]{14pc}\caption{A Lorentz boost deforms the spatial slice
    $t={\rm constant}$ in a linear fashion, the slope of the new slice
    determined by $v/c$. \label{Lorentz}}
\end{minipage}
\end{figure}

Space-time structure can be encoded algebraically and geometrically. An
observer moving at constant speed $v$ follows the trajectory $x(t)=x_0+vt$.
According to special relativity, we assign new coordinates
\[
 x'=\frac{x-vt}{\sqrt{1-v^2/c^2}}\quad,\quad
 ct'=\frac{ct-vx/c}{\sqrt{1-v^2/c^2}}
\]
to events as seen by the moving observer. A spatial slice $t={\rm constant}$
is transformed to a slanted slice $t'={\rm constant}$ obeying the equation
$ct=(v/c)x+{\rm constant}$ in the old coordinates; Fig.~\ref{Lorentz}.

\begin{figure}[h]
\includegraphics[width=12pc]{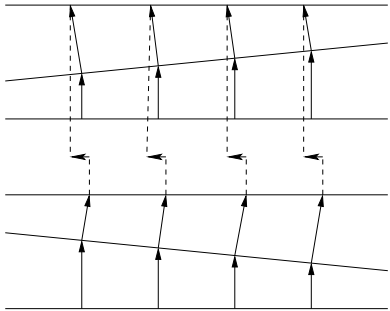}\hspace{2pc}%
\begin{minipage}[b]{20pc}\caption{Commutator of a boost and a time
    translation. In the top part, we first perform a boost by $v$, then wait
    some time $\Delta t$ and undo the boost; in the bottom part, the first
    boost is by $-v$, undone after time $\Delta t$. The image points after
    both deformations lie on the same spatial slice in both orderings,
    but are displaced by $\Delta x =v\Delta t$: During the time interval
    $\Delta t$, all objects have moved with velocity $v$. \label{HypDefLin}}
\end{minipage}
\end{figure}

Just as boosts, all Poincar\'e transformations can be realized geometrically
as linear deformations of spatial slices, with functions $N(\vec{x})=c\Delta
t+(\vec{v}/c)\cdot \vec{x}$ implementing time translations and boosts by
normal deformations, and spatial vector fields $\vec{w}(\vec{x})= \Delta
\vec{x}+\bf R\vec{x}$ for spatial translations and rotations within a spatial
slice. Also their commutators, that is the whole Poincar\'e algebra, can be
geometrized. As an example, we look at the commutator of boosts and time
translations: In Fig.~\ref{HypDefLin}, we see a normal deformation by
$N_1(x)=v x/c$ (a boost with velocity $v$) followed by a normal deformation by
$N_2(x)=c\Delta t-v x/c$ (the reverse Lorentz boost, and waiting $\Delta t$),
performed in the two possible orderings. As elementary geometry shows, a boost
of velocity $v$ commutes with a time translation by $\Delta t$ up to a spatial
displacement $w(x)=\Delta x=v\Delta t$.

\begin{figure}[h]
\includegraphics[width=14pc]{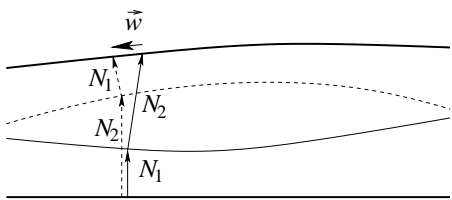}\hspace{2pc}%
\begin{minipage}[b]{20pc}\caption{Non-linear hypersurface
    deformations. Repeated deformations along normals, in different orderings,
    differ by spatial shifts $\vec{w}$. Geometry implies the commutator
    (\ref{TT}). (See e.g.\ \cite{Action} for a detailed
    calculation.) \label{SurfaceDef}}
\end{minipage}
\end{figure}

We have geometrized uniform, inertial motion. In order to extend these
constructions to general relativity, we must implement local Lorentz
transformations: linear deformations with locally changing slope. Our
transformations become arbitrary, non-linear coordinate changes ---
geometrically, we consider non-linear deformations of space. There are now
infinitely many generators, given by spatial deformations $S(\vec{w}(x))$
along (shift) vector fields $\vec{w}$, and normal timelike deformations
$T(N(x))$ by (lapse) functions $N(x)$. Geometry in diagrams such as
Fig.~\ref{SurfaceDef}, generalizing Fig.~\ref{HypDefLin}, produces the
hypersurface-deformation algebra
\begin{eqnarray}
 [S(\vec{w}_1),S(\vec{w}_2)]&=& S({\cal L}_{\vec{w}_2}\vec{w}_1) \label{SS}\\
{} [T(N),S(\vec{w})] &=& T(\vec{w}\cdot\vec{\nabla}N)\\
{} [T(N_1),T(N_2)] &=& S(N_1\vec{\nabla}N_2-N_2\vec{\nabla}N_1) \label{TT}\,.
\end{eqnarray}
The last equation, via gradients $\vec{\nabla}$, makes use of a spatial metric.

Hypersurface deformations geometrize the symmetries of space-time. Symmetry
often determines dynamics. In the case of hypersurface deformations, a theory
invariant under them is generally covariant \cite{DiracHamGR}. As shown by
\cite{Regained,LagrangianRegained}, second-order field equations for geometry
(a spatial metric tensor), invariant under the hypersurface-deformation algebra
as gauge transformations, must equal Einstein's.  Classically, these are
canonical expressions of the well-known fact that the Einstein--Hilbert
action, as a 2-parameter family with Newton's and the cosmological constant,
is unique among all second-order generally covariant actions
(\ref{HigherCurv}) for the space-time metric. (For more details on canonical
gravity, see \cite{CUP}.)

The geometrized view does not tell us much new about classical gravity. But it
turns out to be useful for quantum gravity: One of the symmetries is a time
translation, generated by the Hamiltonian. The Hamiltonian is of prime
interest for quantized dynamics, and usually receives quantum corrections. If
we quantize gravity and amend the classical Hamiltonian by quantum
corrections, generators of the hypersurface-deformation algebra, and perhaps
even the algebra itself, change. In the geometrized view, the question of
quantum gravity, in very general terms, can be formulated thus: How does
quantum physics change hypersurface deformations?

Loop quantum gravity is perhaps the best-developed canonical approach to
quantum gravity. Not coincidentally, it has produced results about the
hypersurface-deformation algebra. However, these features are often hidden
because they require a proper treatment of the off-shell algebra
(\ref{SS})--(\ref{TT}), without fixing the space-time gauge or
deparameterizing. Only recently have such structures become accessible. 

The theory starts with a classical formulation of gravity more akin to
Yang--Mills theories, describing space-time geometry by su(2)-valued
``electric fields'' $\vec{E}_i$ and ``vector potentials'' $\uvec{A}_i$. The
electric field is a densitized triad, determining spatial distances and angles
by three orthonormal vectors $\vec{E}_i$, $i=1,2,3$, at each point in
space. The vector potential, the Ashtekar--Barbero connection, is a
combination of different measures of curvature of space. (Arrows above the
letter indicate contravariant vectors; arrows underneath, covariant ones. If
the metric is one of the basic fields to be quantized, here written in terms
of $\vec{E}_i$, the tensorial type of fields must be kept track of, for a
metric factor inserted to pull up or down indices could significantly change
operator properties after quantization.)

The great advantage of these fields, compared to the spatial metric and its
rate of change or extrinsic curvature, is that they offer natural smearings,
of $\uvec{A}_i$ along curves (holonomies) and $\vec{E}_i$ over surfaces
(fluxes). Some kinematical divergences --- delta functions in Poisson brackets
--- can be eliminated, allowing one to represent both fields as operators. By
analogy with the harmonic oscillator or Fock-space methods in quantum field
theory, holonomies then play the role of creation operators by which
one can construct the state space, and fluxes, some kind of number operator,
determine the excitation level.

\begin{figure}[h]
\includegraphics[width=20pc]{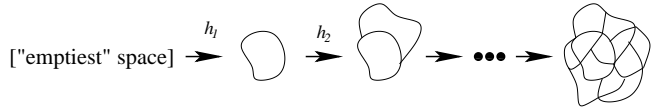}\hspace{2pc}%
\begin{minipage}[b]{14pc}\caption{Holonomy operators generate discrete,
    lattice-like geometry out of ``emptiest'' space. \label{Holonomies}}
\end{minipage}
\end{figure}

In what follows, we use a U(1)-connection $\uvec{A}$ to illustrate salient
features.  Holonomies then read $h_e=\exp(i\int_e\md \lambda
\uvec{A}\cdot\vec{t}_e)$, integrated along curves $e$ in space with tangent
$\vec{t}_e$.  Starting with a simple basic state $\psi_0$, defined by
$\psi_0(\uvec{A})=1$, we construct excited states
\begin{equation}
 \psi_{e_1,k_1;\ldots;e_i,k_i}(\uvec{A})=
\hat{h}_{e_1}^{k_1}\cdots \hat{h}_{e_i}^{k_i}\psi_0(\uvec{A})
 =\prod_{e}
h_e(\uvec{A})^{k_e}=\prod_{e} \exp(ik_e \smallint_e \md \lambda
\uvec{A}\cdot\vec{t}_e)\,.
\end{equation}
Collecting all curves used in a graph $g=\{e_1,\ldots,e_i\}$, we view these
states and their superpositions as discrete lattice-works, or discrete quantum
space, illustrated in Fig.~\ref{Holonomies}. The starting state $\psi_0$ has
no excitations, not even of geometry. It is emptier than the matter vacuum,
emptier than empty space.

Flux operators, constructed from $\vec{E}$ canonically conjugate to
$\uvec{A}$, are derivative operators $\int_S\md^2y
\uvec{n}\cdot\hat{\!\vec{E}}$ integrated over surfaces $S$ in space with
co-normal $\uvec{n}$. They act by
\begin{equation} \label{Flux}
\int_S\md^2y \uvec{n}\cdot\hat{\!\vec{E}}\psi_{g,k}= -8\pi i
G\hbar\int_S\md^2y \uvec{n}\cdot \frac{\delta \psi_{g,k}}{\delta \uvec{A}(y)}=
8\pi \ell_{\rm P}^2\sum_{e\in g} k_e{\rm  Int}(S,e) \psi_{g,k}
\end{equation}
with the intersection number ${\rm Int}(S,e)$ and the Planck length $\ell_{\rm
  P}=\sqrt{G\hbar}$.  These operators have discrete spectra, implying
discrete geometry \cite{AreaVol,Area}: for gravity, fluxes represent the
spatial metric.  The elementary structure of space is realized at the Planck
scale $\ell_{\rm P}\sim 10^{-35}{\rm m}$, at least kinematically, but may be
excited to larger sizes $\sqrt{k}\ell_{\rm P}$ with integer $k$.

It remains challenging to formulate a consistent version of discrete dynamics,
for instance of cosmic expansion, but some general features are known
\cite{QSDI}.  The classical Hamiltonian of gravity depends on $\uvec{A}_i$,
quantized by holonomies. Every one of these operators acts by creating new
geometrical excitations along curves $e$, or new lattice sites in a
discrete picture. The changing excitation level of geometry is measured by
fluxes.

Yang--Mills theory on Minkowski space-time has a Hamiltonian $H_{\rm
  Yang-Mills}=\kappa \int{\rm d}^3x (|\vec{E}_i|^2+|\vec{B}_i|^2)$ for
$\vec{B}_i= \uvec{\nabla}\times\uvec{A}_i+ C_{ijk}\uvec{A}_j\times
\uvec{A}_k$, with structure constants $C_{ijk}$.  Gravity on any space-time
makes use of similar fields, but in a more complicated, non-polynomial way:
\begin{equation} \label{Ham}
 H_{\rm gravity}= \frac{1}{16\pi G}\int{\rm
  d}^3x \frac{\sum_{ijk}\epsilon_{ijk}(\vec{B}_i\times
  \vec{E}_j)\cdot\vec{E}_k}{\sqrt{\frac{1}{6}|\sum_{ijk}\epsilon_{ijk}
(\vec{E}_i\times
    \vec{E}_j)\cdot\vec{E}_k|}} +\cdots
\end{equation}
with $C_{ijk}=\epsilon_{ijk}$.  

This Hamiltonian implies characteristic corrections when quantized in the
loopy way. We have already noted that holonomies are used to quantize
$\uvec{A}_i$ (or $\vec{B}_i$), replacing linear or quadratic functions with
non-polynomial ones. Expanding holonomies $h_e(\uvec{A}_i)$ in Taylor series,
one can match the leading terms with the classical expression, but
higher-order terms remain as corrections motivated by quantum geometry. Flux
operators are linear in $\vec{E}_i$, but they imply corrections nonetheless:
Flux operators (\ref{Flux}) have discrete spectra containing zero and
therefore lack inverse operators, but classically we must divide by a function
of $\vec{E}_i$ to compute the Hamiltonian (\ref{Ham}). It turns out that
Hamiltonian operators can be defined \cite{QSDI,QSDV}, in a more round-about
way circumventing direct inverses of fluxes. Additional quantum
corrections result.

One can illustrate these features by a simpler system: inverse momentum on a
circle with canonical variables $(\varphi,p)$. We have states
$\langle\varphi|n\rangle= \exp(i n\varphi)$ for all integer $n$, a momentum
operator $\hat{p}|n\rangle= \hbar n|n\rangle$, and ``$\hat{p}^{-1}$'' is not
densely defined: it would be infinite on $|0\rangle$.  Instead, we use the
classical identity
\[
 I:=\frac{1}{2}|p|^{-1/2}{\rm sgn}(p)= 
\cos(\varphi)\{\sin(\varphi) ,\sqrt{|p|}\}-
\sin(\varphi)\{\cos(\varphi),\sqrt{|p|}\}
\]
to quantize $\hat{I}=-i\hbar^{-1}
\left(\widehat{\cos}\left[\widehat{\sin},\sqrt{|\hat{p}|}\right]-
  \widehat{\sin}\left[\widehat{\cos},\sqrt{|\hat{p}|}\right]\right)$ as a
well-defined operator with the correct classical limit at large $n$, in
regimes with $\hbar n=p_n$ finite for $\hbar\to 0$.

\begin{figure}[h]
\includegraphics[width=18pc]{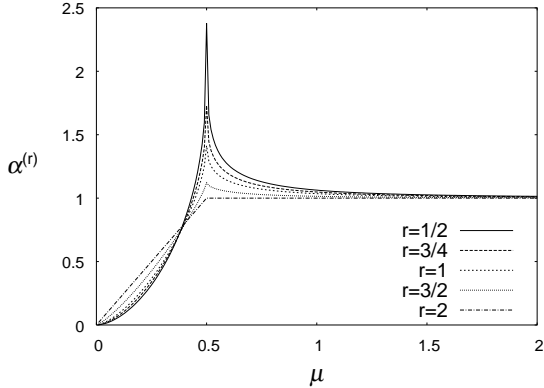}\hspace{2pc}%
\begin{minipage}[b]{14pc}\caption{Inverse-triad corrections
    $\alpha=\langle\hat{I}\rangle(\mu)/I(\mu)$: the ratio of quantum
    expectation value and classical inverse in terms of flux eigenvalues
    $\mu$. (The parameter $r$ is an example for a quantization
    ambiguity.) \label{alpha}}
\end{minipage}
\end{figure}

Applying this procedure to the gravitational Hamiltonian, we have
inverse-triad corrections from quantizing
\[
 \left\{\uvec{A}^i,\int{\sqrt{|\det E|}}\mathrm{d}^3x\right\}= 2\pi G
 \epsilon^{ijk} \frac{\vec{E}_j\times\vec{E}_k}{{\sqrt{|\det E|}}}\:{\rm
   sgn}(\det E)\,.
\]
As illustrated in Fig.~\ref{alpha}, using the calculations for quantizations
and expectation values developed in \cite{InvScale,QuantCorrPert,InflTest}, an
automatic cut-off of $1/E$-divergences at degenerate triads results, but also
corrections for large flux values.  In addition, we have higher-order
corrections from using holonomies for $\vec{B}_i$, and, as always in
interacting quantum theories, quantum back-reaction of fluctuations and higher
moments of a state on expectation values.

All these corrections modify the classical dynamics.
Inverse-triad corrections, for instance, enter the Hamiltonian by
\[
 \frac{1}{16\pi G}\int{\rm
  d}^3x\: \alpha(\vec{E}_l)\: \frac{\sum_{ijk}\epsilon_{ijk}(\vec{B}_i\times
  \vec{E}_j)\cdot\vec{E}_k}{\sqrt{\frac{1}{6}|\sum_{ijk}\epsilon_{ijk}
(\vec{E}_i\times
    \vec{E}_j)\cdot\vec{E}_k|}} +\cdots
\]
The Hamiltonian, in turn, generates time translations as part of the
hypersurface-deformation algebra.  Computing algebraic relations for the
modified Hamiltonian, ensuring that a closed algebra is still realized (the
quantization is then anomaly-free), one finds that the
hypersurface-deformation algebra does change \cite{ConstraintAlgebra}. It is
modified, but not broken:
\begin{eqnarray}
 [S(\vec{w}_1),S(\vec{w}_2)]&=& S({\cal L}_{\vec{w}_2}\vec{w}_1)\\
{} [T_{\alpha}(N),S(\vec{w})] &=& T_{\alpha}(\vec{w}\cdot\vec{\nabla}N)\\
{} [T_{\alpha}(N_1),T_{\alpha}(N_2)] &=& S(\alpha^2(N_1\vec{\nabla}N_2-N_2\vec{\nabla}N_1))\,.
\end{eqnarray}
By inverse-triad corrections we deform but do not violate
covariance. Consistent deformations of the same form have by now been derived
in different models and with various methods, including effective constraints
of spherical symmetry \cite{JR,LTBII,ModCollapse,ModCollapse2} and
operator calculations in $2+1$-dimensional and other models
\cite{TwoPlusOneDef}.

Some immediate consequences can be seen by specializing the algebra to linear
$N$ and $\vec{w}$, comparing with Fig.~\ref{HypDefLin}: The relation between
spatial displacements and boost velocities is modified: $\Delta
x=\alpha^2v\Delta t$, with $\alpha\not=1$ in general. For fluxes larger than
the Planck area, $\alpha>1$ according to Fig.~\ref{alpha}: discrete space
speeds up propagation, much like discrete matter structures can change
propagation speeds of light or phonons.  We are dealing with a new form of
space-time structure, a form which we cannot yet handle directly in the
absence of a manifold picture, perhaps a non-commutative one. Several
subtleties remain in the usual concepts employed to analyze effects of general
relativity, especially the notion of horizons \cite{ModifiedHorizon}. But some
consequences, for instance in cosmological perturbation equations, can be
analyzed based only on the modified $S$ and $T_{\alpha}$: These functions tell
us how fields transform under changes of perspective in space-time --- we are
able to derive gauge-invariant perturbations --- and how they evolve when we
move forward in time using $T_{\alpha}$ \cite{ScalarGaugeInv}.

\begin{figure}[h]
\includegraphics[width=12pc]{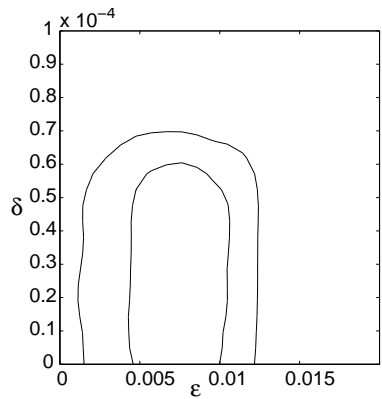}\hspace{2pc}%
\begin{minipage}[b]{14pc}\caption{Likelihood plot for allowed values of the
    slow-roll parameter $\epsilon$ and the inverse-triad correction
    $\delta$. Observations are consistent with $\delta=0$ and provide an
    upper bound $\delta<10^{-4}$ \cite{InflConsist}.\label{s2}}
\end{minipage}
\end{figure}

The modified dynamics of density perturbations $u$ and gravitational waves
$w$, the two gauge-invariant fields of most interest, can be condensed to
\cite{LoopMuk}
\begin{equation}
 -u''+s(\alpha)^2\Delta u +(\tilde{z}''/\tilde{z})u=0\quad,\quad
 -w''+\alpha^2\Delta w +(\tilde{a}''/\tilde{a})w=0
\end{equation}
with functions $\tilde{a}$, $\tilde{z}$ and $s$ all related to the correction
function $\alpha$ derived from inverse-triad operators.  Different modes
propagate with different speeds, giving rise to corrections to the
tensor-to-scalar ratio and other potentially observable effects.  The theory
can be tested, and it is falsifiable: $\delta:=\alpha-1$ is large for small
lattice spacing (see Fig.~\ref{alpha}), while large lattice spacings imply
strong discretization effects and holonomy corrections. One can estimate that
$\delta>10^{-8}$ is required by these theoretical arguments of mutual
consistency \cite{LoopMuk}. Observations give upper bounds on $\delta$: We
have two-sided bounds on the discreteness scale, not just an upper bound that
could always be evaded by tuning parameters. An analysis
\cite{InflTest,InflConsist} using the currently available observational data
provides the upper bound $\delta<10^{-4}$, Fig.~\ref{s2}. While four orders of
magnitude still remain to be bridged to rule out the theory, the bounds are
much closer to each other than the Planck length is to the cosmic Hubble
scale, an immense range in which pure curvature corrections would be
viable. For additional observational results regarding tensor modes, see
\cite{TensorHalfII,TensorSlowRoll,TensorBackground}

Inverse-triad corrections, depending on the microscopic discreteness scale of
a state in relation to the Planck length, are the main correction in
sub-Planckian curvature regimes. Closer to the big bang, where the Planck
density looms, all corrections, including holonomy effects, are
important. Many surprising consequences have been explored in models of loop
quantum cosmology \cite{LivRev,Springer}.

Discreteness means that the wave function of the universe is subject to a
difference equation \cite{Sing,IsoCosmo},
\begin{equation}
C_+(\mu) \psi_{\mu+1}- C_0(\mu)\psi_{\mu}+ C_-(\mu)\psi_{\mu-1} =
\hat{H}_{{\rm matter}}(\mu)\psi_{\mu}
\end{equation}
with wave functions in the triad representation, depending on flux eigenvalues
$\mu\in{\mathbb R}$.  Following the recurrence, a state is extended uniquely
through the classical singularity at $\mu=0$. The singularity is resolved by
this mechanism of quantum hyperbolicity \cite{BSCG}. In the deep quantum
regime, holonomy corrections and dynamical quantum effects are strong and it
is difficult to find an intuitive geometrical picture for the resolution
mechanism. Fortunately, some properties can be seen in solvable, harmonic
models \cite{BouncePert}: the state bounces, reaching a non-zero minimal
volume close to the Planck density \cite{QuantumBigBang}. This consequence is
realized in a special model in which no quantum back-reaction occurs, a flat
isotropic model with a free massless scalar. Approximately, the same behavior
is realized for models with curvature or small matter masses and
self-interactions, but the scenario remains to be confirmed under more general
situations \cite{QuantumBounce,BounceSqueezed}.

\begin{figure}[h]
\includegraphics[width=14pc]{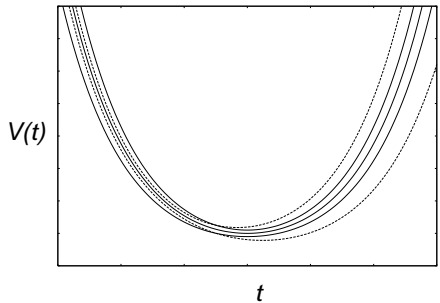}\hspace{2pc}%
\begin{minipage}[b]{14pc}\caption{Dynamical wave functions illustrated by
    their expectation values (central lines) and fluctuations (spreads). The
    expectation value always bounces at some minimal volume $V$ in this
    harmonic holonomy-modified model, but fluctuations on the two sides can
    differ in significant ways. \label{EffBounce}}
\end{minipage}
\end{figure}

Even though the singularity is resolved, some features demonstrate that
space-time, if it even exists in a meaningful form in those extremes, behaves
in surprising ways. One of these features is cosmic forgetfulness
\cite{BeforeBB,Harmonic}, addressing the question of how much pre-big bang
information can be recovered after a state has moved through a resolved
classical singularity. Fig.~\ref{EffBounce} shows sample solutions, the spread
around the central lines indicating quantum fluctuations. Different
fluctuations can be realized before and after the high-density phase,
quantitatively expressed by a squeezing parameter of states
\cite{BounceCohStates}.

In more detail, the ratio of volume fluctuations $\Delta V$ at late and early
times $t$ is bounded by
\begin{equation}
\left|1-\frac{\Delta V_{t\to-\infty}} {\Delta V_{t\to\infty}}\right|
\stackrel{\displaystyle<}{\sim} \frac{\Delta H_{\rm
      matter}/\langle\hat{H}_{\rm
      matter}\rangle}{(\Delta
      V/\langle\hat{V}\rangle)_{t\to\infty}}\,,
\end{equation}
derived with semiclassical techniques and dynamical coherent states
\cite{BounceCohStates,Harmonic,LoopScattering}.  The right-hand side of this
inequality is typically large: the validity of quantum field theory on curved
space-time at $t\to\infty$ indicates that matter-energy fluctuations are much
larger than geometry fluctuations.  The ratio $\Delta V_{t\to-\infty}/{\Delta
  V_{t\to\infty}}$, and evolutionary aspects influenced by it, remains largely
unconstrained, even if the classical singularity can be resolved.

Space-time at and before the big bang is not necessarily classical, even if
isotropic models remain well-defined. Additional information about space-time
can be gained from the hypersurface-deformation algebra, now evaluated with
holonomy corrections for strong-curvature regimes. Preliminary results in
different models \cite{JR,ThreeDeform,ScalarHol,Action} (taking into account
partial holonomy effects) indicate that the algebra is again deformed:
\begin{eqnarray}
 [S(\vec{w}_1),S(\vec{w}_2)]&=& S({\cal L}_{\vec{w}_2}\vec{w}_1)\\
{} [T_{\beta}(N),S(\vec{w})] &=& T_{\beta}(\vec{w}\cdot\vec{\nabla}N)\\
{} [T_{\beta}(N_1),T_{\beta}(N_2)] &=& S(\beta(N_1\vec{\nabla}N_2-N_2\vec{\nabla}N_1))
\end{eqnarray}
with a function $\beta$ depending on $\uvec{A}_i$, or on curvature.  In
contrast to inverse-triad corrections, holonomy corrections can result in
$\beta<0$, which they do at high density (the ``bounce'').  In constructions
such as Fig.~\ref{HypDefLin}, a velocity in one direction would imply motion
in the opposite direction, rather intransigent behavior. More meaningfully,
one can interpret the phenomenon as signature change \cite{Action}: Euclidean
4-dimensional space has a hypersurface-deformation algebra with
$\beta=-1$. What appears as a ``bounce'' in simple isotropic models is not a
bounce at all; there is no evolution and no fully deterministic relation
between pre- and post-big bang states.

Loop quantum gravity implies radical modifications at Planckian densities,
with Euclidean space instead of space-time.  There is no propagation of
structure from collapse to expansion, as assumed in bounce models.  Instead,
loop quantum cosmology shows a non-singular beginning of the Lorentzian phase,
when the diluting density falls sufficiently below the Planck density. Such a
place, nonsingular and yet unprecedented, is natural for posing initial
conditions, for instance for an inflaton state.  We obtain a mixture of linear
and cyclic models.  On less extreme scales, loop quantum gravity implies
space-time, but with modified structures, allowing quantum-gravity corrections
more significant than higher-curvature terms.  Inverse-triad corrections are
not extremely far from testability; loop quantum gravity is falsifiable.

\medskip

\noindent This work was supported in part by NSF grant PHY0748336.

\section*{References:}

\smallskip


\end{document}